# Super-resolved FRET imaging by confocal fluorescence-lifetime single-molecule localization microscopy

*Cecilia Zaza[1†], Germán Chiarelli[2†], Ludovit P. Zweifel[3], Mauricio Pilo-Pais[2], Evangelos Sisamakis[4], Fabio Barachati[4], Fernando D. Stefani[1,5*] and Guillermo P. Acuna[2*]*

[1]Centro de Investigaciones en Bionanociencias (CIBION), Consejo Nacional de Investigaciones Científicas y Técnicas (CONICET), Godoy Cruz 2390, C1425FQD Ciudad Autónoma de Buenos Aires, Argentina.

[2]Department of Physics, University of Fribourg, Chemin du Musée 3, Fribourg CH-1700, Switzerland.

[3]Biozentrum, University of Basel, Spitalstrasse 41, CH-4056 Basel, Switzerland

[4]PicoQuant GmbH, Rudower Chaussee 29 (IGZ), 12489 Berlin, Germany.

[5]Departamento de Física, Facultad de Ciencias Exactas y Naturales, Universidad de Buenos Aires, Güiraldes 2620, C1428EHA Ciudad Autónoma de Buenos Aires, Argentina.

[†] both authors contributed equally

**Corresponding Authors**

* E-mail: fernando.stefani@df.uba.ar (F.D.S.), guillermo.acuna@unifr.ch (G.P.A.)

**Abstract**

FRET-based approaches are a unique tool for sensing the immediate surroundings and interactions of (bio)molecules. FRET imaging and FLIM (Fluorescence Lifetime Imaging Microscopy) enable the visualization of the spatial distribution of molecular interactions and functional states. However, conventional FLIM and FRET imaging provide average information over an ensemble of molecules within a diffraction-limited volume, which limits the spatial information, accuracy, and dynamic range of the observed signals. Here, we demonstrate an approach to obtain super-resolved FRET imaging based on single-molecule localization microscopy using an early prototype of a commercial time-resolved confocal microscope. DNA Points Accumulation for Imaging in Nanoscale Topography (DNA-PAINT) with fluorogenic probes provides a suitable combination of background reduction and blinking kinetics compatible with the scanning speed of usual confocal microscopes. A single laser is used to excite the donor, a broad detection band is employed to retrieve both donor and acceptor emission, and FRET events are detected from lifetime information.

1. **Introduction**

Fluorescence Resonance Energy Transfer (FRET) from a fluorophore in the excited state (donor, *D*) to a fluorophore in the ground state (acceptor, *A*) provides a signal sensitive to the *D-A* separation distance in the $2 - 10$ nm range, the local refractive index, and the spectral overlap between *D* emission and *A* absorption, which in turn depends on various environmental parameters. Based on this, a vast library of imaging protocols and sensors based on FRET have been developed and successfully applied to reveal the spatial distribution, states of association and function of biomolecules in the cellular context.[1,2]

Fluorescence Lifetime Imaging (FLIM) also provides information about local molecular environment through specific probes, for example sensitive to pH or polarity.[3–7] Also, lifetime measurements provide information about energy transfer processes to other fluorophores (FRET),[8,9] metallic nanoparticles (NSET),[10,11] metallic films (MIET),[12] or graphene (GIET).[13–15] A limitation of conventional FLIM or FRET imaging lies in the fact that they provide average information for an ensemble of molecules or nanostructures included within a diffraction-limited volume.

On the other hand, super-resolution microscopy, also known as far-field fluorescence nanoscopy, has revolutionized biological imaging because it maintains the low invasiveness and specificity of fluorescence imaging and at the same time delivers deep sub-diffraction resolution,[16] reaching the sub-10 nm resolution regime.[17] Therefore, the combination of super-resolution imaging with FLIM/FRET holds unique advantages. For example, it could be applied to monitor biomolecular processes taking place in separate but adjacent sub-diffraction biological structures, as recently shown through intensity-based STED-FRET measurements,[18] or to visualize the nanoscale distribution of energy transfer or exciton diffusion events in other material systems such as light-emitting devices,[19] conducting polymers,[20] or nanostructured semiconductors.[21] In addition, the spatial averaging of the FLIM/FRET signal would be significantly reduced, thereby increasing the accuracy and dynamic range of the local functional measurement. Remarkably, while these benefits were clear upon the onset of super-resolution microscopes,[22] and although FRET has been used to enhance or supplement super-resolution methods,[23–27] developing super-resolution FLIM/FRET methods has been challenging.[28]

Conventional Single Molecule Localization Microscopy (SMLM) relies on highly sensitive multi-pixel detectors (e.g. sCMOS or EM-CCD) whose time resolution is not suitable for fluorescence lifetime measurements, although recent technological developments look promising for the near future.[29] Coordinate-targeted (scanning) super-resolution methods can be implemented using pulsed lasers and fast detectors (such as APDs and PMTs), compatible with FLIM. However, these methods usually rely on modulating the emitting fluorescent state, like in STED, which complicates a reliable determination of fluorescence lifetimes. Despite these obstacles, there have been different implementations of simultaneous lifetime measurements and super-resolution imaging. The group of Krachmalnicoff has synchronized an EM-CCD and APDs in a custom-built configuration to obtain both super-resolved positioning and lifetime determinations of single molecules.[30,31] This approach implies a compromise between spatial and temporal resolutions since the limited fluorescence photons detectable from the single molecules must be divided into the two types of detectors. Furthermore, each APD has an associated reduced imaging area in order to avoid the crosstalk in the lifetime determination of each individual fluorophore. Thiele *et al*. have recently performed FLIM measurements on a custom-built fast scanning confocal microscope.[32] A galvo-scanner enabled the acquisition of 10 μm × 10 μm images at a scan rate of 27 Hz. DNA-PAINT and d-STORM were employed to super-resolve cell samples. For the DNA-PAINT experiments, the background signal from the imager strands in solution was strongly reduced by using a low concentration (under 1 nM) in combination with highly labeled samples. The authors used the lifetime information exclusively to identify different fluorophores (Alexa 647 and ATTO 655) operating in similar spectral ranges. Also recently, new techniques such as pulsed interleaved MINFLUX (p-MINFLUX) and RASTMIN allow for super-resolution imaging with clean lifetime information at the expense of a reduced field of view and lower throughput.[33,34]

Here, we show that super-resolved FRET images can be obtained by performing SMLM with an early prototype of a commercial single photon counting confocal microscope (Luminosa, PicoQuant GmbH) using a single laser. For the proof-of-concept demonstration, we exploit the unique advantages of the DNA origami technique in terms of nanometer positioning and stoichiometric control. We show that FRET events separated by sub-diffraction distances can be distinguished based on lifetime modifications.[35] We discuss how the different acquisition parameters affect the final resolution and show that measurements on sparsely labeled samples can be performed using a self-quenched imager at high concentration.[36] Since this approach is based on hardware widely available in numerous labs, it should facilitate the widespread application of super-resolved FLIM/FRET.

## 2. Results and discussion

Super-resolved FLIM was achieved by performing SMLM measurements with a commercial time-resolved, laser-scanning confocal microscope (early prototype of Luminosa microscope, PicoQuant) as schematized in **Figure 1a**. Excitation was done with a pulsed diode laser emitting at 532 nm with pulse width of 76 ps and repetition rate of 40 MHz (LDH-D-FA-530L). The excitation light was spatially filtered with a polarization-maintaining single-mode fiber. A $\lambda/4$ wave-plate was used to create circular polarization. Images were acquired with a laser scanner based on oscillating mirrors (FLIMBee, PicoQuant). A single-photon avalanche diode (SPAD; Excelitas AQRH-14) and suitable optical filters were used to detect fluorescence photons in the spectral range between 545 nm and 728 nm. Photon detection events were time-correlated to excitation pulses using a time-correlated single-photon counting (TCSPC) device (MultiHarp 150 8P, PicoQuant).

A sketch of the DNA origami employed to demonstrate super-resolved FRET imaging is shown in Figure 1b (further details can be found in Table S1 and Figure S1 of the Supplementary Information).[37] With a length of 180 nm, the origami includes three sites for DNA-PAINT, separated by 65 nm. Sites 1 and 2 have 12 and 10 docking strands for DNA-PAINT, respectively. Site 3 features 5 docking strands for DNA-PAINT and 4 fixed ATTO 647N fluorophores designed to act as acceptors. Each of the 5 docking strands of site 3 is located within FRET range to two different ATTO 647N fluorophores (see zoom of site 3 in Figure 1b, *D-A* separation distances range from 5 to 6.3 nm). The DNA origami design also includes 6 biotinylated ssDNA staples on the underside of the structure for sample immobilization onto glass coverslips functionalized with neutravidin.

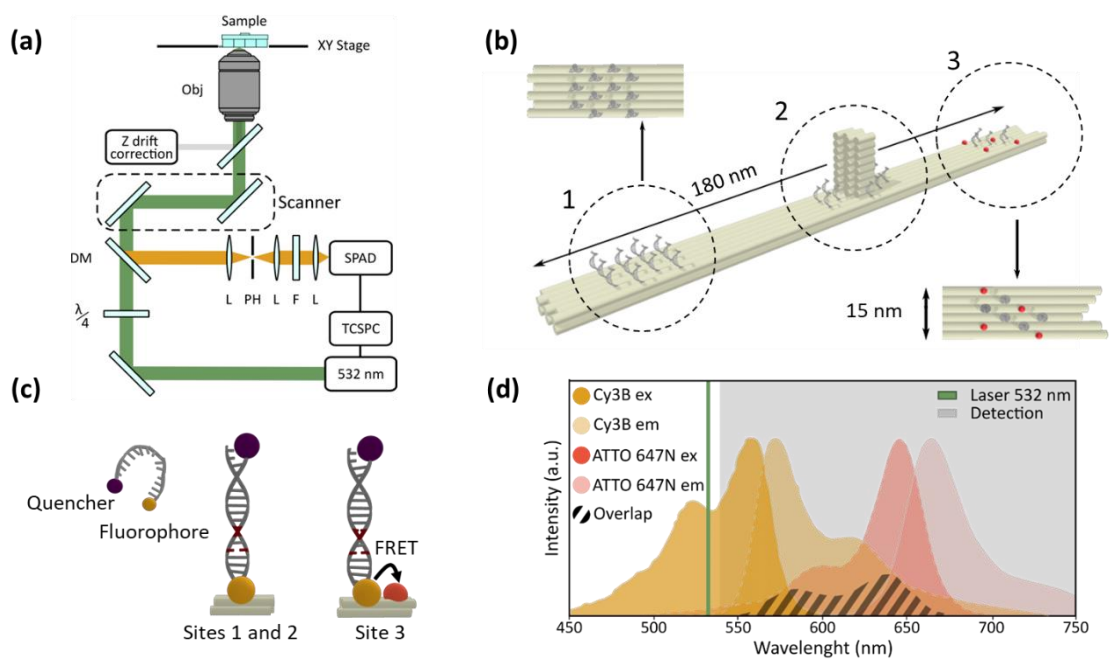

**Figure 1**: Experimental set-up and DNA origami sample. (a) Scheme of the experimental set-up. 532 nm, pulsed laser: λ/4: quarter lambda retardation plate; DM: dichroic mirror; Obj: objective lens; PH: pinhole; F: long-pass filter; L: lens; SPAD: single photon avalanche photodiode; TCSPC: time-correlated single-

photon counting device. (b) Schematic of the DNA origami used for the experiments with three sites with docking strands for DNA-PAINT (1, 2, and 3). Site 3 includes 4 ATTO 647N fixed fluorophores (red dots) that can act as FRET acceptors when a ssDNA imager with a Cy3B dye binds to any docking strand of site 3. (c) Schematic of the fluorogenic imager probe used in solution (left) and when hybridized to the docking strand (right). At site 3, FRET is expected to occur to the nearby acceptors. Mismatches in the binding sequence are represented as unconnected red lines in the dsDNA schematic. (d) Spectral design of the FRET experiment showing the emission and absorption of the donor (*D*, Cy3B) and acceptor (*A*, ATTO 647N), their spectral overlap for FRET, the excitation laser line and the detection window.

A fluorogenic probe, based on the previous work of Chung *et al.*,[36] was used to reduce the typical background signal of DNA-PAINT measurements (Figure 1c). These imager strands differ from those frequently used for DNA-PAINT in that, besides a fluorophore in one end (Cy3B at the 3´end), a quencher is added at the other end (BHQ2 at the 5´end). Also, these imager strands are longer than most implementations of DNA-PAINT with a total of 15 nucleotides. The reason for this is two-fold. First, the strands need to be sufficiently long to enable their folding, decreasing the fluorophore-quencher separation. Second, upon hybridization of the imager strand with the complimentary docking strand, the high number of nucleotides also guarantees that the fluorophore-quencher separation will be increased and the fluorescence signal recovered due to a reduced interaction with the quencher. To accelerate the unbinding kinetics, 3 mismatched bases are included to the imager strands as shown in red in Figure 1c. Details of the sequence used for the fluorogenic imager and docking sites can be found in Table S2 of the Supplementary Information. Figure 1d shows the spectral design of FRET experiments using Cy3b and ATTO 647N as *D* and *A*, respectively.

Samples also included 60 nm gold nanoparticles as fiducial markers to monitor and correct for sample drift. A full description of the sample preparation can be found in the Supplementary Information (Figure S2 in particular for more details on drift correction with gold nanoparticles).

The proper folding and functionality of the DNA origami was verified by performing conventional two-color DNA-PAINT super-resolution measurements in a wide-field microscope (described in Figure S3 of the Supplementary Information). The parameters used for video acquisition were: 10k frames, 100 ms per frame, 130 nm pixel size, a field of view of 30 μm$^2$ and 15 nM of fluorogenic imager in Trolox Glox oxygen scavenger buffer was employed.[38] Excitation was done at 532 nm, and two detection channels were set-up on the same camera chip to detect $D$ and $A$ fluorescence. In **Figure 2a**, a superposition of both channels is presented color-coded: localizations detected in the $D$ channel are shown in green, whereas localizations detected in the $A$ channel are shown in red. As expected, in the $D$ channel each DNA origami exhibits 3 sites, whereas in the $A$ channel only one site is detected at one end of the origami, in agreement with our design (see Figure 1b). At 532 nm, direct excitation of the $A$ (ATTO 647N) is negligible and thus the $A$ localizations are generated by FRET.

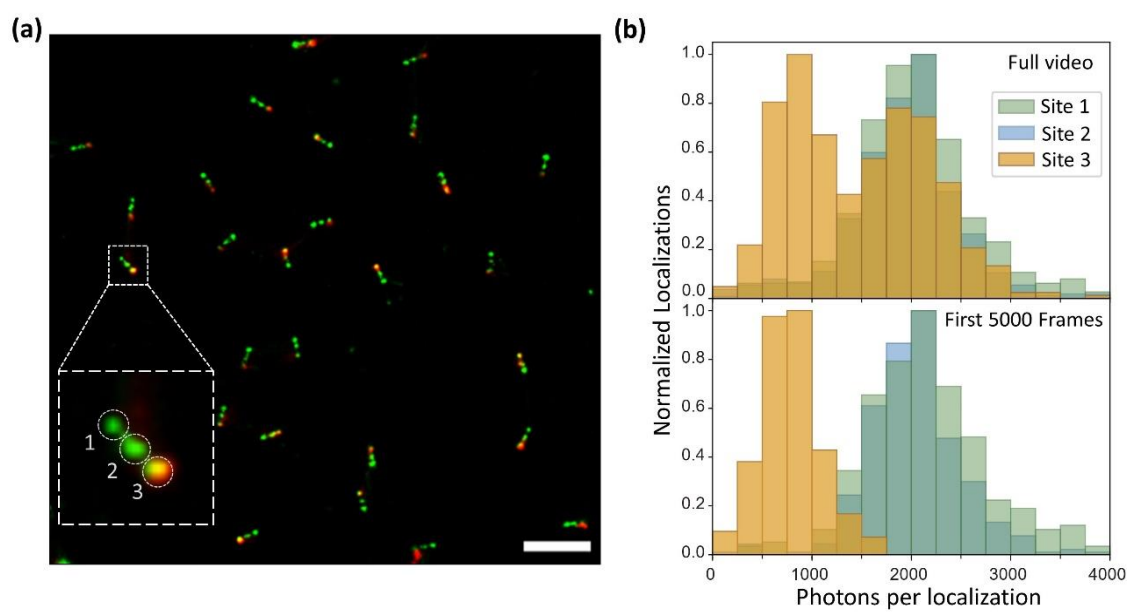

**Figure 2:** Sample characterization through wide-field measurements. (a) Two-color super-resolution image

of the origami sample. Scale bar: 500 nm. The three different binding sites are identified using the acceptor emission in the red channel (highlighted in the zoom). (b) Photon histograms on the $D$ channel for each localization for a single DNA origami, classified by binding site.

The occurrence of FRET at site 3 was further verified by studying the $D$ photon counts of the single molecule binding events on each site, as shown in Figure 2b. The average photon count of $D$ emission in site 3 (defined as the one showing $A$ emission; see inset in figure 2a) was found to be consistently lower than for the other two sites. Also, for measurement times long enough to allow for significant photobleaching of the acceptors, site 3 presents a bimodal distribution of photon counts, with a second peak with the same average as sites 1 and 2. The FRET efficiency can be estimated by comparing the photon counts of the $D$ in the presence and absence of $A$.[39,40] A value of $0.6 \pm 0.1$ was obtained, in fair agreement with calculations based on the estimated Förster radius and the $D - A$ distances extracted from the DNA origami design (further details on the Supplementary Information).

Once confirmed that the DNA origami samples exhibit DNA-PAINT spots with and without FRET separated by sub-diffraction distances, we set out to perform super-resolved FRET imaging based on FLIM using a scanning confocal microscope. For this type of measurements, the scanning area, number of pixels and the pixel dwell time, define the measuring time per image. These parameters need to be adjusted to the binding and unbinding kinetics of the imager strands mostly defined by the imager concentration and number of complimentary nucleotides, respectively. In particular, the time needed to acquire an image must be significantly smaller than the average duration of the single-molecule emission events, which is mainly determined by the unbinding kinetics but could be reduced due to photobleaching of the fluorophores depending on the laser power and other experimental conditions. We employed a 15 nM concentration of imager strands, with 12 complimentary bases to the docking strands. This configuration resulted in single molecule emission events with a typical duration of $1 - 2$ s. The scanning parameters were set to image a field of view (FoV) of $7 \times 7$ μm$^2$, using $60 \times 60$ pixels with a pixel dwell time of 10 μs. Data was acquired only in one scanning direction and with a dead-time of $0.8 - 1.5$ ms between images. Under these conditions, the average collection time per image was 73 ms, so that the raw data consisted of a series of confocal images (video) acquired at 13.6 Hz. Prior to further analysis, the raw data was binned by combining 10 successive confocal images to create what will be hereafter referred to as "frames for SMLM". This binning was implemented in order to increase the signal to background and to minimize distortion effects associated with blinking of single molecules during scans. Then, each frame corresponds to an acquisition time of roughly 700 ms, which approximately matches the imager's binding time. Finally, the binned video was analyzed as it is commonly performed in SMLM measurements using Picasso.[41]

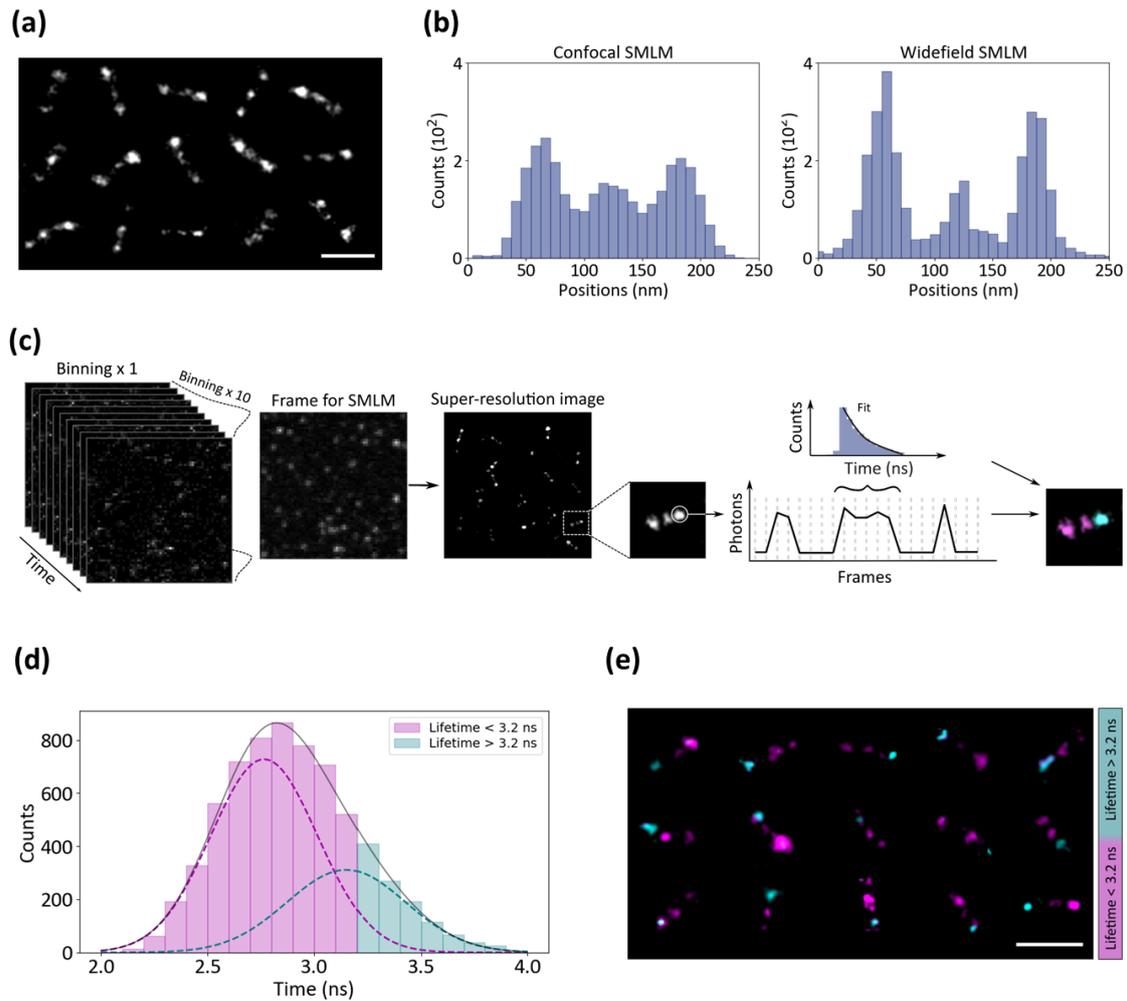

**Figure 3:** Super-resolution FRET measurements based on FLIM. (a) Examples of super-resolution imaging of DNA origami samples obtained through DNA-PAINT measurements on a confocal microscope. (b) Histogram plot of the localizations obtained with the confocal (left) and the widefield (right) microscopes. For the comparison, the same area size and number of localizations were employed. (c) Schematic of the measurement pipeline for super-resolved FLIM. Raw confocal images are binned to obtain a frame for single-molecule localization microscopy. Next, for each single molecule emission event used for a localization, photon arrival times are analysed to obtain the excited state lifetime. (d) Distribution of lifetimes of all localizations in a particular FoV. The total distributions of lifetimes is well represented by a sum of two Gaussian distributions (black line), one ascribed to the *D* in the absence of FRET (magenta) and the other to *A* and *D* undergoing FRET (cyan). A threshold of 3.2 ns was used to differentiate single-molecule emission presenting FRET (lifetime > 3.2 ns) or no FRET (lifetime < 3.2 ns). (e) Super-resolved FLIM of a selection of DNA origami, using a binary lifetime color code according to (d), and with the intensity proportional to the number of localizations. Scale bars 200 nm.

**Figure 3a** shows a confocal SMLM image reconstructed from a series of 10k frames. The three DNA-PAINT sites of the DNA origami are clearly resolved. Figure 3b shows the localization histogram of 4 origamis extracted from Figure 3a (left) aligned using Picasso Average module, along with a comparison obtained using widefield microscopy (right) in a similar area and comparable number of localizations per origami. These findings demonstrate that super-resolution imaging based on SMLM DNA-PAINT can be performed using commercial scanning confocal microscopes. The achieved resolution of 20 nm was lower than the 10 nm achieved with widefield camera based SMLM measurements. This can be attributed to the larger drift experienced during image acquisition as the overall measurement time for widefield camera based SMLM was x7 times shorter which makes drift correction for confocal SMLM in the postprocessing analysis more challenging particularly considering that the early prototype of the microscope employed was not yet optimized to minimize drift.

Next, we analyzed the data to retrieve the photon arrival times of each single-molecule emission event used to reconstruct the super-resolution SMLM image, as schematically shown in Figure 3c. The fluorescence lifetime of each single-molecule emission event was determined by fitting the histogram of all photon detection times associated with that event with a single-exponential, similarly to the work by Thiele et al.[32] Figure 3d shows the distribution of single-molecule lifetimes for a measurement of 7000 localizations, which spans from 2.2 to 4 ns. The total distribution of lifetimes is well described by a sum of two Gaussians populations (black full line). The population with shorter lifetime is centered at 2.7 ns and ascribed to donor molecules that do not undergo FRET (magenta dotted line)[42]. The longer lifetime population (cyan dotted line) represents the weighted average lifetime of the $A$ (3.85 ns for ATTO 647N)[26] and the $D$ lifetimes, considering the average FRET efficiency of 0.6. A lifetime threshold can be used to discriminate single-molecule emission events involving FRET from events not showing FRET. Setting for example, a lifetime threshold of 3.2 ns (Figure 3d), classifies 83% of the events above the threshold with FRET. Figure 3e shows the FLIM super-resolution images of 15 DNA origamis, where lifetimes above and below a threshold value of 3.2 ns are color-coded as in Figure 3d. The occurrence of FRET at site 3 was super-resolved for each DNA origami.

3. **Conclusion**

In conclusion, super-resolution FRET imaging can be performed with a commercial time-resolved confocal microscope, using a single laser excitation tuned to the donor absorption, and a single detection channel covering both donor and acceptor emission. The method introduced in this work enables the combination of SMLM with lifetime measurements. Scanning speed and acquisition must be matched to the single molecule blinking kinetics. We found that a fast scanning and a subsequent binning of images is convenient to obtain complete images of the blinking single molecules, necessary to perform precise localizations. The experiment design must meet some key conditions. First, the excitation wavelength should produce negligible acceptor excitation. Second, the acceptor lifetime should be significantly different, preferentially longer than the donor lifetime. In this way, FRET events are detected through acceptor emission, which in turn is discerned from donor emission based on lifetime measurements. In our experiments, SMLM was achieved by DNA-PAINT using fluorogenic imager strands which were advantageous to reduce background. Fiducial nanoparticles were used to correct for sample drift during post-processing analysis, which becomes relevant due to the relatively longer measurement times. We foresee that the presented methodology may be useful for numerous groups as it enables super-resolution FLIM and FRET imaging using conventional equipment. Further, improving confocal microscopes with active correction for lateral and axial sample drift would lead to FLIM and FRET imaging with spatial resolutions equivalent to camera-based SMLM.

**Supporting Information**

Supporting Information is available from the Wiley Online Library or from the author.

**Acknowledgements.**

F.D.S. acknowledges the support of the Max Planck Society and the Alexander von Humboldt Foundation and the Agencia Nacional de Promoción Científica y Tecnológica (ANPCYT), projects PICT-2017-0870. G.P.A. acknowledges the support of the Swiss National Science Foundation through project 200021_184687 (G.P.A.) and the National Center of Competence in Research Bio-Inspired Materials NCCR, project 51NF40_182881. F.B. and E.S. acknowledge the support from the Federal Ministry of Education and Research (BMBF), program "Photonics Research Germany" (project 13N15324).

**Conflict of Interest.**

Fabio Barachati and Evangelos Sisamakis are employees of PicoQuant GmbH.

# References

[1]   W. R. Algar, N. Hildebrandt, S. S. Vogel, I. L. Medintz, *Nat Methods* **2019**, *16*, 815.

[2]     Á. Szabó, T. Szendi-Szatmári, J. Szöllősi, P. Nagy, *Methods Appl Fluoresc* **2020**, *8*, 032003.

[3]     T. W. J. Gadella Jr, T. M. Jovin, R. M. Clegg, *Biophys Chem* **1993**, *48*, 221.

[4]     E. P. Buurman, R. Sanders, A. Draaijer, H. C. Gerritsen, J. J. F. Van Veen, P. M. Houpt, Y. K. Levine, *Scanning* **1992**, *14*, 155.

[5]     M. J. Cole, J. Siegel, S. E. D. Webb, R. Jones, K. Dowling, M. J. Dayel, D. Parsons-Karavassilis, P. M. W. French, M. J. Lever, L. O. D. Sucharov, *J Microsc* **2001**, *203*, 246.

[6]     K. Suhling, L. M. Hirvonen, J. A. Levitt, P.-H. Chung, C. Tregidgo, A. Le Marois, D. A. Rusakov, K. Zheng, S. Ameer-Beg, S. Poland, S. Coelho, R. Henderson, N. Krstajic, *Medical Photonics* **2015**, *27*, 3.

[7]     T. Niehörster, A. Löschberger, I. Gregor, B. Krämer, H.-J. Rahn, M. Patting, F. Koberling, J. Enderlein, M. Sauer, *Nat Methods* **2016**, *13*, 257.

[8]     J. R. Lakowicz, Ed. , Springer US, Boston, MA, **2006**, pp. 1–26.

[9]     G. Sánchez-Mosteiro, E. M. H. P. van Dijk, J. Hernando, M. Heilemann, P. Tinnefeld, M. Sauer, F. Koberlin, M. Patting, M. Wahl, R. Erdmann, N. F. van Hulst, M. F. García-Parajó, *J Phys Chem B* **2006**, *110*, 26349.

[10]    T. L. Jennings, M. P. Singh, G. F. Strouse, *J Am Chem Soc* **2006**, *128*, 5462.

[11]    G. P. Acuna, M. Bucher, I. H. Stein, C. Steinhauer, A. Kuzyk, P. Holzmeister, R. Schreiber, A. Moroz, F. D. Stefani, T. Liedl, F. C. Simmel, P. Tinnefeld, *ACS Nano* **2012**, *6*, 3189.

[12]    A. I. Chizhik, J. Rother, I. Gregor, A. Janshoff, J. Enderlein, *Nat Photonics* **2014**, *8*, 124.

[13]    I. Kamińska, J. Bohlen, R. Yaadav, P. Schüler, M. Raab, T. Schröder, J. Zähringer, K. Zielonka, S. Krause, P. Tinnefeld, *Advanced Materials* **2021**, *33*, 2101099.

[14]    A. Ghosh, A. Sharma, A. I. Chizhik, S. Isbaner, D. Ruhlandt, R. Tsukanov, I. Gregor, N. Karedla, J. Enderlein, *Nat Photonics* **2019**, *13*, 860.

[15]    I. Kaminska, J. Bohlen, S. Rocchetti, F. Selbach, G. P. Acuna, P. Tinnefeld, *Nano Lett* **2019**, *19*, 4257.

[16]    S. W. Hell, S. J. Sahl, M. Bates, X. Zhuang, R. Heintzmann, M. J. Booth, J. Bewersdorf, G. Shtengel, H. Hess, P. Tinnefeld, A. Honigmann, S. Jakobs, I. Testa, L. Cognet, B. Lounis, H. Ewers, S. J. Davis, C. Eggeling, D. Klenerman, K. I. Willig, G. Vicidomini, M. Castello, A. Diaspro, T. Cordes, *J Phys D Appl Phys* **2015**, *48*, 443001.

[17]    L. A. Masullo, A. M. Szalai, L. F. Lopez, F. D. Stefani, *Biophys Rev* **2021**, *13*, 1101.

[18]    A. M. Szalai, B. Siarry, J. Lukin, S. Giusti, N. Unsain, A. Cáceres, F. Steiner, P. Tinnefeld, D. Refojo, T. M. Jovin, F. D. Stefani, *Nano Lett* **2021**, *21*, 2296.

[19] R. P. Puchert, F. Steiner, G. Plechinger, F. J. Hofmann, I. Caspers, J. Kirschner, P. Nagler, A. Chernikov, C. Schüller, T. Korn, J. Vogelsang, S. Bange, J. M. Lupton, *Nat Nanotechnol* **2017**, *12*, 637.

[20] K. Becker, P. G. Lagoudakis, G. Gaefke, S. Höger, J. M. Lupton, *Angew Chem Int Ed Engl* **2007**, *46*, 3450.

[21] J. Liu, L. Guillemeney, B. Abécassis, L. Coolen, *Nano Lett* **2020**, *20*, 3465.

[22] H. E. Grecco, P. J. Verveer, *ChemPhysChem* **2011**, *12*, 484.

[23] S. Cho, J. Jang, C. Song, H. Lee, P. Ganesan, T.-Y. Y. Yoon, M. W. Kim, M. C. Choi, H. Ihee, W. Do Heo, Y. Park, *Sci Rep* **2013**, *3*, 1208.

[24] A. Auer, M. T. Strauss, T. Schlichthaerle, R. Jungmann, *Nano Lett* **2017**, *17*, 6428.

[25] J. Lee, S. Park, W. Kang, S. Hohng, *Mol Brain* **2017**, *10*, 63.

[26] N. S. Deußner-Helfmann, A. Auer, M. T. Strauss, S. Malkusch, M. S. Dietz, H.-D. Barth, R. Jungmann, M. Heilemann, *Nano Lett* **2018**, *18*, 4626.

[27] R. J. Stöhr, R. Kolesov, K. Xia, R. Reuter, J. Meijer, G. Logvenov, J. Wrachtrup, *ACS Nano* **2012**, *6*, 9175.

[28] A. M. Szalai, C. Zaza, F. D. Stefani, *Nanoscale* **2021**, *13*, 18421.

[29] N. Oleksiievets, J. C. Thiele, A. Weber, I. Gregor, O. Nevskyi, S. Isbaner, R. Tsukanov, J. Enderlein, *Journal of Physical Chemistry A* **2020**, *124*, 3494.

[30] D. Bouchet, J. Scholler, G. Blanquer, Y. De Wilde, I. Izeddin, V. Krachmalnicoff, *Optica* **2019**, *6*, 135.

[31] G. Blanquer, B. van Dam, A. Gulinatti, G. Acconcia, Y. De Wilde, I. Izeddin, V. Krachmalnicoff, *ACS Photonics* **2020**, DOI 10.1021/acsphotonics.9b01317.

[32] J. C. Thiele, D. A. Helmerich, N. Oleksiievets, R. Tsukanov, E. Butkevich, M. Sauer, O. Nevskyi, J. Enderlein, *ACS Nano* **2020**, *14*, DOI 10.1021/acsnano.0c07322.

[33] L. A. Masullo, F. Steiner, J. Zähringer, L. F. Lopez, J. Bohlen, L. Richter, F. Cole, P. Tinnefeld, F. D. Stefani, *Nano Lett* **2020**, *21*, 840.

[34] L. A. Masullo, A. M. Szalai, L. F. Lopez, M. Pilo-Pais, G. P. Acuna, F. D. Stefani, *Light Sci Appl* **2022**, *11*, 199.

[35] P. W. K. Rothemund, *Nature* **2006**, *440*, 297.

[36] K. K. H. Chung, Z. Zhang, P. Kidd, Y. Zhang, N. D. Williams, B. Rollins, Y. Yang, C. Lin, D. Baddeley, J. Bewersdorf, *Nat Methods* **2022**, DOI 10.1038/s41592-022-01464-9.

[37] A. K. Adamczyk, T. A. P. M. Huijben, M. Sison, A. Di Luca, G. Chiarelli, S. Vanni, S. Brasselet, K. I. Mortensen, F. D. Stefani, M. Pilo-Pais, G. P. Acuna, *ACS Nano* **2022**, *16*, 16924.

[38]   T. Cordes, J. Vogelsang, P. Tinnefeld, *J Am Chem Soc* **2009**, *131*, 5018.

[39]   J. Enderlein, *Int J Mol Sci* **2012**, *13*, 15227.

[40]   J. Bohlen, Á. Cuartero-González, E. Pibiri, D. Ruhlandt, A. I. Fernández-Domínguez, P. Tinnefeld, G. P. Acuna, *Nanoscale* **2019**, *11*, 7674.

[41]   J. Schnitzbauer, M. T. Strauss, T. Schlichthaerle, F. Schueder, R. Jungmann, *Nat Protoc* **2017**, *12*, 1198.

[42]   N. Oleksiievets, Y. Sargsyan, J. C. Thiele, N. Mougios, S. Sograte-Idrissi, O. Nevskyi, I. Gregor, F. Opazo, S. Thoms, J. Enderlein, *Commun Biol* **2022**, *5*, 1.

Table of Contents

Cecilia Zaza[1,†], Germán Chiarelli[2,†], Ludovit P. Zweifel[3], Mauricio Pilo-Pais[2], Evangelos Sisamakis[4], Fabio Barachati[4], Fernando D. Stefani[1,5*] and Guillermo P. Acuna[2*]

**Super-resolved FRET imaging by confocal fluorescence-lifetime single-molecule localization microscopy**

We present a method for obtaining super-resolved FRET imaging based on single-molecule localization microscopy using the DNA-PAINT technique in combination with fluorogenic probes in a commercial time-resolved confocal microscope. We show that FRET events separated by sub-diffraction distances can be distinguished based on lifetime modifications.

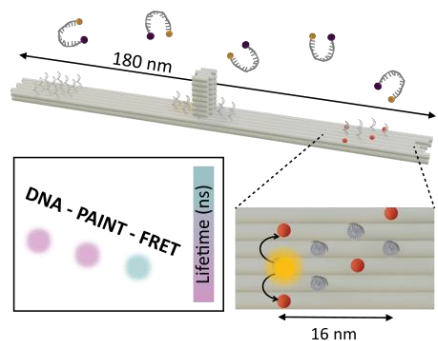